\documentstyle[preprint,aps]{revtex}
\tightenlines
\input epsf
\title{Scattering phases in quantum dots: an analysis based on lattice models}

\author{A. Levy Yeyati$^1$  and M. B\"uttiker$^2$}

\address{$^1$Departamento de F\'{\i}sica Te\'orica de la Materia
Condensada CV, Universidad Aut\'onoma de Madrid, E28049 Madrid, Spain
\\
$^2$D\'epartement de Physique Th\'eorique,
Universit\'e de Gen\`eve, CH-1211, Gen\`eve 4, Switzerland}
\begin{document}

\draft
\maketitle
\begin{abstract}
The properties of scattering phases in quantum dots are analyzed with
the help of lattice models. We first derive the expressions relating the
different scattering phases and the dot Green functions. We
analyze in detail the Friedel sum rule and discuss the deviation of the
phase of the transmission amplitude from the Friedel phase at the zeroes
of the transmission. The occurrence of such zeroes is related to the
parity of the isolated dot levels. A statistical analysis of the
isolated dot wave-functions reveals the absence of significant
correlations in the parity for large disorder and the appearance, for
weak disorder, of certain dot states which are strongly coupled to the
leads. It is shown that large differences in the coupling to the leads
give rise to an anomalous charging of the dot levels. A mechanism for
the phase lapse observed experimentally based on this property 
is discussed and illustrated with model calculations.
\end{abstract}
\vspace{0.3cm}

\vspace{0.3cm}

\narrowtext

\section{Introduction}

Phase coherence is at the heart of most phenomena studied in mesoscopic
physics. However, the behavior of the electronic wave-function 
phase {\it itself} in an actual quantum transport device had not been 
studied until recent years. 
In the case of quantum dots, investigations have been predominantly
restricted to
conductance measurements \cite{review} 
which carry no information on the transmission
phase. It was not until the experiments by Yacoby et al.
\cite{Yacoby} that the interest in phase behavior of quantum dots 
really started.  
In these experiments a quantum dot was embedded in one of the arms
of an Aharonov-Bohm (AB) ring in an attempt to analyze the transmission
phase evolution as a function of the dot gate voltage. Although these
experiments were the first to demonstrate the presence of a coherent
component in the current through a quantum dot in the Coulomb blockade 
regime, they failed to give the complete evolution of the phase. This
limitation was explained \cite{us} as a consequence of the phase locking
(or phase rigidity) that occurs in a two terminal geometry. It was shown
\cite{us} that the AB effect in such geometry is characterized by a {\it
parity}: as a function of the AB flux the conductance exhibits either a
local maximum at zero flux (positive parity) or a local minimum
(negative parity). Another intriguing feature of the experimental
results was the ``parity conservation'' over a large sequence of Coulomb
blockade peaks, which reflected a similar evolution of the phase over
each peak. The complete evolution of the phase could be obtained in a
subsequent experiment by Schuster et al. \cite{Schuster} using a four
terminal geometry. This experiment confirmed the expected evolution of
the phase around the peaks and revealed that an abrupt jump of $\pi$
occurs in the valleys between the peaks. 

Since 1995 several theoretical efforts have been devoted to explain
these observations 
\cite{us,Bruder,HW1,Yacoby2,Deo,Gefen,HW2,HW3,Xu,Ryu,Baltin,Kang,Silverstrov,BG}: 
Ref. \cite{us} proposed a screening effect, Ref. \cite{Bruder}
alluded to dot degeneracies, Refs. \cite{HW1,HW2,HW3} associated the
observed effect to an asymmetric deformation of the dot which leads to
repeated charging of the same dot level and Ref. \cite{Silverstrov}
pointed out some special properties of the dot states in a semi-chaotic
situation (this mechanism will be further analyzed in the present work).
In spite of all these efforts
there is still the feeling that a more fundamental explanation 
is lacking. Each one of the proposed mechanisms can be criticized
as relying on some particular assumptions.
Only the approximate sum rule proposed recently in Ref. \cite{BG} is
supposed to be valid in a generic (chaotic) situation. However, there is 
still no experimental evidence of the near resonance phase-lapse predicted by
this mechanism.

On the other hand, the phase problem affects our knowledge on generic
properties of scattering phases. One fundamental relation, invoked in Ref.
\cite{us} in connection to this problem, is the Friedel sum rule which
relates the phase of the eigenvalues of the scattering matrix to the
charge accumulated in the dot region. Being related to the dot charge,
the Friedel phase is a continuous function of the system parameters and
cannot exhibit an abrupt behavior like the one found in the experiments
by Schuster et al. \cite{Schuster}. However, as pointed out recently by
Lee\cite{Lee} and Taniguchi and one of the authors\cite{Taniguchi}
the phase of the transmission
amplitude can depart from the Friedel phase and exhibit a non-analytic
behavior at the points where the modulus of the transmission vanishes.  
It is thus intersting to study the general conditions for the occurrence
of zeroes in the transmission through a quantum dot. 

The aim of this paper is to investigate the behavior of the different
scattering phases in quantum dots with the help of lattice models. These
type of models allow to describe a dot of arbitrary shape and to study
the influence of disorder \cite{Verges}.
We shall first derive the expressions for the
different scattering phases in terms of Green functions. These expressions
allow to relate the occurrence of zeroes in the transmission
with the parity of the isolated dot wave-functions. We also study the
statistical properties of the dot wave-functions in a disordered 
quantum dot, showing explicitly the absence of significant
correlations in the chaotic case. On the other hand, as suggested in 
Ref. \cite{Silverstrov}, for weak disorder 
one can identify certain dot levels which are much more strongly coupled 
to the leads than average. We shall show that in this situation the dot
levels are populated in an anomalous way as a function of gate voltage.
We shall finally discuss the possible role of this type of correlation
effects in the phase problem.   

The outline of the paper will be the following:
In section II we introduce a generic lattice model for a Quantum Dot
coupled to single-moded leads. In section III we derive the expressions
for the different scattering phases in terms of Green functions. We
discuss in particular the Friedel sum rule and the relation of the
Friedel phase to the phase of the transmission amplitude. In section IV
we study the conditions for the occurrence of zeroes in the
transmission and show that they are independent of the strength of the
coupling to the leads. The statistical
properties of the isolated dot wave-functions are analyzed in section V. 
Finally, in section VI we discuss the role of electron correlation
effects. We end the paper with some conclusions and final remarks.

\section{Generic lattice model}

As a model for a two dimensional quantum dot we consider a collection
of $N$ sites on a square lattice (see Fig. 1). This model can represent
a dot of arbitrary shape. The electrons in the dot are described by a
tight-binding Hamiltonian with site energies $\epsilon_i$ and a constant
hopping element $t$ coupling nearest-neighbors only. ($t$ will be taken
as the unit of energy).  The site energies
are allowed to vary following an imposed electrostatic confining potential
and/or randomly in order to study the influence of disorder.

On the other hand, electron-electron interactions can be included within
the constant charging energy model by adding a term
$V_{coul} = E_C (N_{dot} - C V_g/e)^2$, where $N_{dot}$ is the mean
number of electrons in the dot, to the one-electron Hamiltonian.
Its effect will be discussed in section VI.

We shall consider that the dot is coupled to electron reservoirs
by two one-dimensional leads as depicted in Fig. 1. We may assume that
the coupling to
the left and right leads is restricted to two sites labeled by 1 and
$N$ respectively. The first site on each lead are connected to these two
sites on the dot by hopping elements $t_L$ and $t_R$ respectively. 
As will be discussed later, this situation can be
easily generalized to the case where the first sites on the leads 
are connected to several sites on the dot (this multiple connection
is illustrated in Fig. 1 by the dashed lines). 

\section{Scattering phases and Green functions}

The electronic properties of lattice models are conveniently given
in terms of Green functions. We need to introduce the retarded
and advanced Green operators given by

\begin{equation}
\hat{G}^{r,a}(\omega) = \left[ \omega - \hat{H}_{dot} - 
\hat{\Sigma}^{r,a}(\omega) 
\right]^{-1} ,
\end{equation}

\noindent
where $\hat{H}_{dot}$ is the one-electron part of the isolated dot Hamiltonian
and $\hat{\Sigma}^{r,a}$ is the (retarded, advanced) self-energy operator
describing the coupling of the dot to the leads. In a Fermi-liquid like
description, the self-energy should
also contain terms accounting for electron-electron interactions
\cite{multilevel}.
We shall postpone its discussion to section VI and concentrate on the
one-electron properties of the quantum dot until then. When the coupling 
to the leads is localized at sites 1 and $N$ we have

\begin{equation}
\Sigma^{r,a}_{l,m}(\omega) = \delta_{l,j_{\alpha}} \delta_{m,j_{\alpha}} 
t_{\alpha}^2 g^{r,a}_{\alpha} (\omega) ,
\end{equation}

\noindent
where $\alpha= L,R$, $j_L=1, j_R=N$ and $g^{r,a}_{\alpha}$ are the local 
Green functions at the semi-infinite one-dimensional leads.

In terms of the Green operator one can express the dot total density of
states $\rho(\omega)$ as

\begin{equation}
\rho(\omega) = \frac{1}{2 \pi i} \mbox{Tr}\left[ \hat{G}^a(\omega)
- \hat{G}^r(\omega) \right] = \frac{1}{\pi} \mbox{Im Tr} \left[
\hat{G}^a(\omega)\right] .
\end{equation}

We shall now discuss the connection of this quantity to the scattering
matrix $\hat{S}(\omega)$, defined in terms of Green functions
by means of the generalized Fisher-Lee relations \cite{FisherLee}

\begin{equation}
S_{\alpha, \beta}(\omega) = \delta_{\alpha, \beta} -
2 i \sqrt{\Gamma_{\alpha} \Gamma_{\beta}}
G^r_{j_{\alpha},j_{\beta}}(\omega)  ,
\end{equation}

\noindent
where we have introduced the
tunneling rates $\Gamma_{\alpha} = t^2_{\alpha} \mbox{Im}g^a_{\alpha}$.
The unitarity of $\hat{S}$ can be readily shown (see appendix A).
We start by defining the quantity $\theta_F$ as

\begin{equation}
\theta_F(\omega) = \mbox{Im} \mbox{ln} \mbox{Det}
\left[\omega - \hat{H}_{dot} - \hat{\Sigma}^a(\omega) \right] .
\end{equation}

The derivative with respect to the energy of $\theta_F$ is then given
by

\begin{equation}
\frac{\partial \theta_F}{\partial \omega} = \mbox{Im}
\mbox{Tr} \left[\hat{G}^a  \left( 1 - \frac{\partial \hat{\Sigma}^a}
{\partial \omega} \right) \right]  .
\label{derivativetheta}
\end{equation}

Thus, in the case when the energy dependence of the self-energy can be neglected
one obtains the identity
\begin{equation}
\frac{\partial \theta_F}{\partial \omega} = \pi \rho(\omega). 
\end{equation}

This case corresponds to a particular type of leads, having a large
density of states, which can screen any deviation
from charge neutrality induced by the presence of the dot. This type of
leads can be called ``non-polarizable leads".

On the other hand, it can be show (see appendix B) that $\theta_F$ can be 
expressed in terms of the scattering matrix $\hat{S}$ as

\begin{equation}
\theta_F(\omega) = \frac{1}{2i} \mbox{ln} \mbox{Det}\left[ \hat{S}(\omega) 
\right] 
\end{equation}
and thus one obtains a relation between the dot total density of states
and the derivative of the scattering matrix with respect to the energy

\begin{equation}
\rho(\omega) = \frac{1}{2 \pi i} \frac{\partial}{\partial \omega}
\mbox{ln} \mbox{Det}\left[\hat{S}(\omega)\right] .
\end{equation}

Then, by integrating this expression up to the Fermi energy, we obtain
the generalized Friedel sum rule

\begin{equation}
N_{dot} = \frac{1}{\pi} \theta_F(E_F) .
\label{fsr}
\end{equation}

We see that $\theta_F$ is an important scattering phase. As it is related
to the dot charge it should be a continuous function of the energy. It has
also a simple relation to the eigenvalues of the scattering matrix. Due
to the unitarity of $\hat{S}$ its eingenvalues are of the form 
$e^{2 i \xi_{1,2}}$ and thus $\theta_F = \xi_1 + \xi_2$.

It should be emphasized that relation (\ref{fsr}) holds only for the case of
non-polarizable leads. For a more general case one should include also the
charge induced on the leads in the Friedel sum rule.
The deviation between the dot total density of states and the derivative
of the Friedel phase with respect to energy has also been pointed out by 
Gasparian et al. \cite{Gasparian} who analized the connection between 
densities of states and the scattering matrix for continuous models. 
In particular Eq. (14) in \cite{Gasparian} can be written as

\begin{equation}
\frac{\partial \theta_F}{\partial \omega} = \pi \rho(\omega) - \mbox{Im}
\left(\frac{s_{L,L} + s_{R,R}}{4 \omega}\right) 
\end{equation}
which coincides with our Eq. (\ref{derivativetheta}) provided that 
we make the approximation

\begin{equation}
\frac{\partial \hat{\Sigma}}{\partial \omega} \approx \frac{i \mbox{Im} 
\hat{\Sigma}}{2 \omega} .     
\end{equation}

Another scattering phase which is relevant for the interference
phenomena observed in the experiments is the phase of the transmission
amplitude $\theta_t = \arg S_{LR}$.
In some particular
cases (for instance in a one-dimensional problem) $\theta_F$ and $\theta_t$
coincide. However, as noticed recently by some authors \cite{Lee,Taniguchi}, 
they are in general different. While $\theta_F$ is a continuous function,
$\theta_t$ may not be defined at certain energies where the transmission
vanishes. In order to be more precise, one can parametrize
a general scattering matrix as

\begin{equation}
\hat{S} = \left( \begin{array}{cc} i e^{i(\theta+\varphi_1)} \sin \phi
&  e^{i(\theta+\varphi_2)} \cos \phi \\
e^{i(\theta-\varphi_2)} \cos \phi & i e^{i(\theta-\varphi_1)} \sin \phi
\end{array} \right) .
\end{equation}

\noindent
with real phases $\theta$, $\varphi_1$, $\varphi_2$ and $\phi$. It is then
easy to show that $\theta_F = \theta + \pi/2$. On the other hand, when time
reversal symmetry holds one has $S_{LR}=S_{RL}$ and thus $\varphi_2=0$,
in which case the argument of the transmission amplitude is related to
$\theta_F$ by

\begin{equation}
\theta_t = \theta_F + \pi \Theta(\cos{\phi}) - \frac{\pi}{2} ,
\end{equation}

\noindent
where $\Theta(x)$ is the step function. Therefore, $\theta_t$ exhibits
jumps of $\pi$ each time $\cos{\phi}$ changes sign, i.e. at the points
where there is a zero of the transmission. At these points the phase of
the transmission amplitude deviates from the Friedel phase.
Notice that the abrupt jump
of $\pi$ of the phase of the AB oscillations between consecutive resonances
is a central feature of the experimental results of Schuster et al. 
\cite{Schuster}. We thus conclude that the study of the 
occurrence of zeroes of the transmission
is essential to understand the experimentally observed behavior.

\section{Conditions for zeroes of the transmission}

Within lattice models one can establish precise conditions for the occurrence
of zeroes in the transmission amplitude through the dot. According to
the Fisher-Lee relations the condition for having a zero in $S_{LR}$
at energy $E_0$
is that $G^r_{1N}(E_0) = 0$. This is gives 

\begin{equation}
G^r_{1N}(E_0) = \frac{C_{1N}(E_0 - \hat{H}_{dot} - \hat{\Sigma}^r(E_0))}
{\mbox{Det} (E_0 - \hat{H}_{dot} - \hat{\Sigma}^r(E_0))} = 0 ,
\label{cond1}
\end{equation}

\noindent
where $C_{ij}(\hat{A})$ denotes the cofactor of the element $i,j$ in 
the matrix $\hat{A}$.
It is easy to see that the polynomial in the numerator is real and does
not depend on the self-energy coupling of the dot to the leads at sites
1 and $N$. This is a direct consequence of having the coupling to the
leads localized at these sites. We can thus reduce the condition for
the zeroes, Eq. (\ref{cond1}), to the simpler expression

\begin{equation}
C_{1N}\left(E_0 - \hat{H}_{dot}\right) = 0 .
\label{cond2}
\end{equation}

\noindent
Eq. (\ref{cond2}) clearly shows that the zeroes of the transmission 
are characteristic
of the isolated dot structure and do not depend on the strength of the
coupling to the leads. This is illustrated in Fig. 2 where
the transmission for a 5x5 sites dot is shown for varying values of $\Gamma$
within an energy range having a zero. One can observe that while the shape
of the transmission varies substantially, the position of the zero is not
affected.

This property allows one to relate the zeroes of the transmission to the
wavefunctions of the isolated dot. In the weak coupling limit one can
approximate $G_{1N}$ as

\begin{equation}
G^a_{1N} \approx 
\sum_n \frac{\psi^n_1 \psi^n_N}{\omega - \lambda_n - i \Gamma_L
(\psi^n_1)^2 - i \Gamma_R (\psi^n_N)^2} ,
\end{equation}

\noindent
where $\lambda_n$ and $\psi^n_j$ denote the eigenvalues and the amplitudes
of the corresponding wavefunction for the isolated dot. The condition to
have a zero between two consecutive eigenvalues $\lambda_n, \lambda_{n+1}$
is then simply given by $\psi^n_1 \psi^n_N \psi^{n+1}_1 \psi^{n+1}_N > 0$.

We can now identify the sign of $\psi^n_1 \psi^n_N$ as the {\it parity}
of the corresponding dot wavefunction. By this reasoning we conclude 
that there should be a zero of the transmission in between dot states
with the same parity {\it regardless of the strength of the coupling to
the leads}.

In real systems inelastic scattering would prevent the occurrence of
exact zeroes in the transmission. This situation can be described within
our model by additional leads coupled to the dot as voltage probes
\cite{MB2}. This effect is discussed in the next subsection.

\subsection*{A simple example}

A simple example which already exhibits a zero in the transmission amplitude 
is the case of a four sites sites dot, i.e. $N=4$. In this model sites 
1 and 4 are the
ones coupled to the leads. Sites 2 and 3 are coupled to sites 1 and 4 by
hopping elements $t$ and we take the site energies on 1 and 4 as 
$\epsilon_1 = \epsilon_4 = 0$. The transmission amplitude for such model 
is given by 

\begin{equation}
S_{LR} = -2 i \sqrt{\Gamma_L \Gamma_R} \frac{2 t^2 (\omega -
\bar{\epsilon})}{(\omega + i \Gamma_L)(\omega + i \Gamma_R)
(\omega - \epsilon_2)(\omega - \epsilon_3) - 4t^2 
(\omega - \bar{\epsilon})(\omega + i \bar{\Gamma})}  ,
\label{SLR}    
\end{equation}

\noindent
where $\bar{\epsilon} = (\epsilon_2 + \epsilon_3)/2$ and $\bar{\Gamma} =
(\Gamma_L + \Gamma_R)/2$. 

For weak coupling and $|\epsilon_2 - \epsilon_3| \gg t$ the transmission
has two well resolved resonances at $\omega \simeq \epsilon_{2,3}$. 
The modulus and the phase of $S_{LR}$ are shown in Fig. 3. As can be
observed the phase of the transmission exhibits a similar evolution
around each resonance. At the point $\omega = \bar{\epsilon}$ there is a
zero of the transmission and its phase exhibits an abrupt jump of $\pi$.
From Eq. \ref{SLR} it is clear that the zero is not
dependent on the strength of the coupling to the leads.

Inelastic scattering can be simulated by additional voltage probes
coupled to sites 2 and 3. Let us denote by $\Gamma_{inel}$ the coupling
to the additional leads. It is easy to see that the zero of the
transmission now moves away from the real energy axis to $\bar{\epsilon}
+ i \Gamma_{inel}$. The jump in the phase of the transmission becomes 
smaller than $\pi$ and it is no
longer abrupt but has a finite width given by $\Gamma_{inel}$. 
The width of the phase jump can thus be taken as a measure of broadening
of the dot levels due to inelastic scattering.

\section{Statistical analysis of the transmission phase}

According to the discussion of the previous section, the occurrence of
zeroes in the transmission (and the associated jump in the transmission
phase) is related to the parity of the isolated dot wavefunctions on
consecutive levels. In a similar way as done by many authors for
analyzing the mesoscopic conductance fluctuations of quantum dots in 
the Coulomb
blockade regime \cite{Jalabert}, the parity should be studied
statistically over many realizations of the dot potential.

In order to search for correlations of the parity of the dot
wavefunctions we have numerically diagonalized dot lattice models up
to $30 \times 30$ sites. The confining potential is assumed to have the
form of an isotropic parabola with curvature $\alpha$.
We have investigated the influence of disorder
and also the influence of different models for coupling the dot to the
leads.     

Figure 4a shows the spacing distribution between consecutive levels for
various values of the disorder strength $W$ and $\alpha = 0.01$ (in
units of $t/a^2$, where $a$ is the lattice spacing).
One can notice that the distribution approaches the
the Wigner distribution \cite{Wigner}

\[ p(s) = \frac{\pi}{2} s \exp\left[-\frac{\pi}{4}s^2\right] \]

\noindent
(plotted as a full line in Fig. 4a) as the disorder strength increases. 
This agreement suggests that our results for $W > 1$ should be well 
described by random matrix theory.

For analyzing correlations in the parity of the wavefunctions we plot
in Fig. 4b the probability to find consecutive levels with the
same parity as a function of level spacing. As can be observed, this
probability is almost constant as a function of level spacing. Although
the probability is slightly larger than 0.5 for weak disorder ($W=0.5$)
it approaches 0.5 as the disorder increases.
This behavior indicates that correlations in the parity are negligible
within this model.

The previous results correspond to the case where the dot is coupled to 
the leads at sites 1 and $N$, and thus the parity of the wavefunctions is 
given by $\arg(\psi^n_1 \psi^n_N)$. This situation can be generalized to
the case where many dot sites are coupled to each lead as indicated by
the dotted lines in Fig. 1. 
We shall assume that both leads are coupled to
the same number $N_{leads}$ of neighboring sites on each side of
the dot. 
In this case the parity will be given by $\arg(\sum_{j_L,j_R} \psi^n_{j_L}
\psi^n_{j_R})$, where $j_{L,R}$ denote the sites coupled to the left and
right lead respectively. It should be emphasized that the 
conditions for the occurrence of zeroes discussed in section IV
remains valid in this case, the first sites on the leads $L$ and $R$ 
playing the role of sites 1 and $N$. 

In Fig. 5 we show the total probability to find consecutive levels with
the same parity as a function of $N_{leads}$
and different values of the disorder strength $W$. For $W=0.5$ one
can notice a slight decrease of the probability for increasing
$N_{leads}$. The probability nevertheless remains always close 
to 0.5 indicating the absence of significant parity correlations 
even when changing the model for coupling the dot to the leads. 

These results demonstrate that it is very unlikely to find a large
sequence of dot levels with the same parity and consequently the
difficulty to account for the experimental results of Ref.
\cite{Schuster} within a one-electron model. There are, however, some
peculiarities of the dot states in the limit of weak disorder which can 
give rise to correlation effects as discussed in the next section. These
correlation effects are associated with certain dot states which are 
strongly coupled to the leads as discussed below. 
 
The coupling strength 
$\alpha_n = |\sum_{j_L,j_R} \psi^n_{j_L} \psi^n_{j_R}|$ is shown in
figure 6 as a function of the level number for the case
$N_{leads} = 5$ and different values of $W$. 
In the case of extended contacts to the leads and weak disorder one can
clearly distinguish states which are more strongly coupled to the leads
than the average. For the case $W=0.5$, this states are indicated by
arrows in Fig. 6. One can notice by their level number
(67,79,92,106,..) that they appear in a well defined sequence, 
corresponding to the shell structure of states in the isotropic 2D 
harmonic oscillator. Contour plots of the wavefunctions with level number 
67 and the next three levels are shown in Fig. 7. Although they
correspond to nearly degenerate levels only the first one is strongly
coupled to the leads.   

\section{Effects due to Coulomb interactions}

Up to now we have neglected interactions and concentrated on the
one-electron properties of the dot.
As commented in section II, electron-electron interactions can be
included by means of the constant charging energy model.  
Within the traditional description of Coulomb blockade in quantum dots based
on simple master equations \cite{Beenakker} the main role of the
charging energy is to open a gap between consecutive dot levels, dot
states being populated according to the isolated dot ground state for each
number of electrons. Within this picture, the evolution
of the phase over a sequence of dot resonances will be fixed by the
parity of the corresponding one-electron dot states. As discussed in the
previous section, the parity of one-electron dot states does not exhibit
significant correlations and therefore this picture is unable to account
for the experimental results of Ref. \cite{Schuster}.

The situation may change drastically when taking into account the
finite coupling to the leads beyond lowest order perturbation theory. 
In this case correlation effects may lead to a population of the dot 
levels different to the one predicted by simple master equations. 
This possibility was recently pointed out by Silvestrov and Imry 
\cite{Silverstrov} 
and advanced as an explanation of the experimental 
results on the phase of the transmission through a quantum dot. 
The mechanism proposed by these authors would take place when there is a
strongly coupled dot level followed by nearly degenerate weakly coupled
levels. This situation is characteristic for the case of weak disorder 
discussed in the previous section.  

The basic mechanism proposed by Silverstrov and Imry can be understood by
considering a simple four sites problem with two (spinless) electrons. 
We shall assume 
that two of the sites correspond to the dot levels and the other two 
represent the left and right leads. Let us call $\epsilon_1$ and 
$\epsilon_2$ the one-electron energies of the isolated dot levels and 
$\Delta = \epsilon_2 - \epsilon_1 > 0$ the spacing between them. 
The lower level is symmetrically coupled to the leads by effective hopping 
elements $t_{1L}=t_{1R}=T$  
which are much larger than the ones corresponding to the upper 
level $t_{2L} = t_{2R} = t$. The energy levels on the leads sites are 
taken as zero. Let us analyze how the dot levels are populated as the
gate voltage shift down the energy levels from $\epsilon_1, \epsilon_2 \gg 0$
to $\epsilon_1, \epsilon_2 \ll 0$. 

Finding the ground state of this problem requires in principle to
diagonalize the $6 \times 6$ matrix corresponding to the system
Hamiltonian in the basis ${|n_1 n_2 n_L n_R>}$ where $(n_1,n_2)$ are the
occupation numbers for the dot levels and $(n_L,n_R)$ correspond to the
leads. The problem can, however, be reduced by a basis change in which we
replace the states with one electron in the leads by its symmetric and
anti-symmetric combination, i.e $|1010>$, $|1001>$, $|0110>$ and
$|0101>$ are replaced by $(|1010>\pm|1001>)/\sqrt{2}$ and
$(|0110>\pm|0101>)/\sqrt{2}$. In this way, the initial $6 \times 6$
matrix is reduced into two $3 \times 3$ blocks: one block corresponds
to the empty dot states which couples to the anti-symmetric combinations
and the other to the doubly occupied dot state coupled to the symmetric
combinations, having the form

\[ H^{s} = \left( \begin{array}{ccc} \epsilon_1 + \epsilon_2 + E_c &
\sqrt{2}T & \sqrt{2}t \\ \sqrt{2}t & \epsilon_2 & 0 \\
\sqrt{2}t & 0 & \epsilon_1 \end{array} \right) \;\;\;
H^{a} = \left( \begin{array}{ccc} \epsilon_1  &
0  & \sqrt{2}T  \\ 0 & \epsilon_2 & \sqrt{2}t \\
\sqrt{2}T & \sqrt{2}t & 0  \end{array} \right) \]

In the limit $\Delta \rightarrow 0$, i.e. $\epsilon_2 \rightarrow
\epsilon_1 = -V_g$ the ground state for each symmetry
is given by

\[ \lambda^s = \frac{-3V_g + E_c}{2} - \sqrt{\left(\frac{-V_g+E_c}{2}
\right)^2 + 2 ( T^2 + t^2)}\]
\[ \lambda^a = \frac{V_g}{2} - \sqrt{\left(\frac{V_g}{2}
\right)^2 + 2 ( T^2 + t^2)} \]

When $V_g \ll 0$ the system starts in the anstisymmetric ground
state and the charge of the dot levels evolve with $V_g$ according
to

\[ <n_1> = \frac{2T^2}{(V_g + \lambda^a)^2 + 2(T^2 + t^2)} \;\;\;
<n_2> = \frac{2t^2}{(V_g + \lambda^a)^2 + 2(T^2 + t^2)} \] 
which shows that the dot levels start populating when $-V_g \sim T$.
However, as $t \ll T$, the charge in the weakly coupled level will be
negligible as compared to the strongly coupled one. 

At $V_g = E_c/2$ there is a crossing between $\lambda^a$ and
$\lambda^s$. As a result the symmetric ground state becomes more stable
and the charge on each dot level is now given by

\[ <n_1> = \frac{2t^2 + (V_g + \lambda^s)^2}{(V_g + 
\lambda^s)^2 + 2(T^2 + t^2)} ,\;\;\;
<n_2> = \frac{2T^2 + (V_g  + \lambda^s)^2}{(V_g + 
\lambda^s)^2 + 2(T^2 + t^2)} \]

Thus, when crossing $V_g = E_c/2$ the charge on the strongly coupled
state goes from $\sim T^2/(T^2 + t^2)$ to $\sim t^2/(T^2+ t^2)$ while the
charge on the weakly coupled level does the opposite evolution (we 
neglect small corrections of order $(T/E_c)^2$). 
We see that nearly  
one electron is transferred from the strongly coupled level to the
weakly coupled one in the middle of the charging curve, the total
charge in the dot remaining constant (up to order $(T/E_c)^2$).  
The strongly coupled level is again filled when $V_g$ is close 
to $E_c$.  
The situation is similar for finite $\Delta \ll T$, the
abrupt jump is also present but shifted to $V_g \sim E_c/2 + \Delta$. 
 
A more realistic description of the actual situation requires the
inclusion of many sites to represent both the leads and the dot levels.
One would need a large number of
sites in order to simulate a continuous density of states on the leads.
Using Lanczos method we are able to numerically diagonalize interacting
systems with
up to 16 sites and 8 electrons. The dot levels charge as a function of
gate voltage for a systems with 6 sites on a line for representing each 
lead and 4 sites for representing the dot levels, is shown in Fig. 8. The
dot levels are coupled symmetrically to the leads by hopping parameters
$t_1 = 0.1$ and $t_2 = t_3 = t_4 = 0.02$ (all energies in units of the
charging energy $E_c$) while the hopping parameter within the leads is
taken as $0.1$. The weakly coupled levels lie
at $0.025$, $0.030$, and $0.035$ respectively above the strongly coupled
level. As can be observed, the strongly coupled level is repeatedly
charged and discharged in a similar way as for the simple four sites
model. One can notice, however, that the jumps in the charge of the
strongly coupled level are not so much abrupt and become
progressively less pronounced.  

The next question is how  
this particular charging of
the dot levels affects the transmission amplitude and its phase.
In order to obtain the transmission through the interacting dot we map this 
finite size cluster into an
effective non-interacting system with the same charge on each dot level 
for each value of the gate voltage.
In this effective system we use the same hopping parameters as in the
cluster calculation but replace the leads by infinite one-dimensional
chains. The effective dot levels are determined self-consistently 
to get the correct level charges. The phase of the transmission
amplitude and the transmission probability thus obtained are shown in Fig. 9.
As expected, the phase follows closely the charge of the strongly
coupled level. It should be remarked that this behavior is not dependent
on the parity of the weakly coupled levels.
On the other hand, the transmission probability exhibits  
broad peaks that can be associated with the repeated charging of the
strongly coupled level together with narrow satellite peaks
corresponding to the charging of the weakly coupled levels. These narrow
peaks lie closer to the broader peaks as the cycle number increases. In
the fourth cycle the narrow peak can hardly be resolved.    

\section{Conclusions and final remarks}

To summarize, we have analyzed the behavior of scattering phases in 
quantum dots using lattice models. We have first studied the definition 
of the different scattering phases in terms of Green functions.
We have shown that abrupt jumps of $\pi$ in the phase of the
transmission amplitude are associated
with the occurrence of transmission zeroes.
Then,  we have shown that within lattice models and assuming single moded
leads the zeroes are independent of the strength of the coupling to the
leads \cite{comment}. 
This property allows us to relate the occurrence of a zero with
the parity of consecutive dot states in the isolated dot.

We have studied the statistical properties of the isolated dot
states as a function of the disorder strength. This analysis reveals 
that there are no significant parity
correlations between consecutive dot levels.  
For moderate disorder there appear states which are much more strongly
coupled to the leads than the average. We have studied charging effects
in this situation using exact diagonalization of small clusters. 
We have shown that the dot
resonances as a function of gate voltage may correspond to the strongly
coupled level through several cycles. 

The viability of this mechanism as a possible explanation of the ``phase
problem'' deserves further discussion. The presence of strongly
coupled levels followed by many weakly coupled levels is a reminiscent
of the integrable case which gradually disappears for increasing
disorder (see Fig. 6). In this sense, the mechanism is not universal as
it would not hold for a fully chaotic quantum dot. Thus, this mechanism
could be tested experimentally by leading the dot into a fully chaotic
situation (using, for instance additional gates to distort the dot
shape, as in Ref. \cite{Folk}).
On the other hand, the transmission as a function of gate voltage
exhibits narrow satellite peaks close to the broader Coulomb blockade
peaks (see Fig. 9) which are not observed experimentally. Being so
narrow these peaks could be easily washed out by thermal broadening or
due to the finite bias voltage used in the measurements. 
Of course if certain narrow peaks are not resolved in the experiment,
a number of additional scenarios might account for parity conservation,
even within a non-interacting picture. Differing parity conserving
scenarios can then still be distinguished according to their phase 
evolution pattern.
It is very unlikely that a non-interacting theory can generate a
a phase evolution like that of Fig. 9 which resembles closely that 
seen in the experiments.

One should also comment about the role of spin, not included in the
present analysis. At very low temperatures spin degeneracy would lead to
a Kondo effect. This effect has been recently observed experimentally 
\cite{Kondo} but in somewhat smaller quantum dots and would be reflected
in the phase behavior as a plateau at $\sim \pi/2$ in between a pair of
resonances corresponding to the same dot level \cite{Kang,vonDelft}.
At temperatures larger than the Kondo temperature spin degeneracy could
at most account for the similar phase behavior over two consecutive dot
resonances. Other spin effects, like Hund's rule require almost perfect
symmetry leading to orbital degeneracy and so far have been observed 
only in ultrasmall quantum dots \cite{Tarucha}.

To conclude, 
we point out that our analysis has been restricted
to the case of single-moded leads. This is an assumption which has been
implicitly been used in all the theoretical models of the problem 
following the characterization of the experimental set up given in 
Ref. \cite{Schuster}. Nevertheless, the analysis of what would be
measured in a multichannel situation is of interest and deserves further 
theoretical investigation.

\acknowledgements
The authors would like to thank T. Taniguchi for fruitful discussions.
One of us (ALY) also thanks discussions with F. Flores and J.A.
Verg\'es. 
This work has been partially supported by the Spanish CICyT under
contract PB97-0044 and the Swiss National Science Foundation.

\vspace{2cm}

\appendix{\bf Appendix A}

We show in this appendix the unitarity of the scattering matrix defined
from the Fisher-Lee relations. One has

\begin{eqnarray}
\left(\hat{S}^{\dagger} \hat{S}\right)_{11} & = & |1 - 2 i \Gamma_L G^r_{11}|^2
+ 4 \Gamma_L \Gamma_R |G^r_{1N}|^2  \nonumber \\
\left(\hat{S}^{\dagger} \hat{S}\right)_{12} & = & 
-2 i \sqrt{\Gamma_L \Gamma_R}
\left[ (1 + 2 i \Gamma_L G^a_{11})G^r_{1N} - G^a_{1N} (1 - 2 i \Gamma_R
G^r_{NN}) \right] ,
\end{eqnarray}

\noindent
and similar expressions for $\left(\hat{S}^{\dagger} \hat{S}\right)_{21}$
and $\left(\hat{S}^{\dagger} \hat{S}\right)_{22}$. These elements can
be rewritten as

\begin{eqnarray}
\left(\hat{S}^{\dagger} \hat{S}\right)_{11} & = & 2 i \Gamma_L
\left[ G^a_{11} - G^r_{11} - 2 i \Gamma_L |G^a_{11}|^2 - 2 i \Gamma_R
|G^a_{1N}|^2 \right] + 1 \nonumber \\
\left(\hat{S}^{\dagger} \hat{S}\right)_{12} & = & 
-2 i \sqrt{\Gamma_L \Gamma_R}
\left[ G^r_{1N} - G^a_{1N}  + 2 i \Gamma_L G^a_{11}G^r_{1N} + 2 i \Gamma_R
G^a_{1N}G^r_{NN} \right] .
\end{eqnarray}

It is now easy to show that the expressions between brackets vanish
identically. This fact is a consequence of the identity

\begin{equation}
\hat{G}^a - \hat{G}^r = \hat{G}^a \left[\hat{\Sigma}^a - \hat{\Sigma}^r
\right] \hat{G}^r
\end{equation}

\noindent
which follows from the definition of the Green operator. Taking the
matrix elements (1,1) and (1,N) one finds

\begin{eqnarray}
G^a_{11} - G^r_{11} & = & 2i \Gamma_L |G^a_{11}|^2 + 2i \Gamma_R
|G^a_{1N}|^2 \nonumber \\
G^a_{1N} - G^r_{1N} & = & 2i \Gamma_L G^a_{11} G^r_{1N} + 2i \Gamma_R
G^a_{1N}G^r_{NN} \nonumber \\
\end{eqnarray}

\vspace{2cm}

\appendix{\bf Appendix B} 

We show in this appendix that $2 i \theta_F = \mbox{ln} \mbox{Det}
\left[S\right]$. First notice that

\begin{equation}
\theta_F = \mbox{Im} \mbox{ln} \mbox{Det} \left[ \omega - \hat{H}_{dot} -
\Sigma^a \right] = \frac{1}{2i} \mbox{ln} \mbox{Det} \left[(\omega - 
\hat{H}_{dot} - \Sigma^a) \hat{G}^r \right] .
\end{equation}

From the definition of the Green operator one can rewrite this expression
as

\begin{equation}
\theta_F = \frac{1}{2i} \mbox{ln} \mbox{Det} \left[1 + (\Sigma^r - \Sigma^a)
G^r \right] .
\end{equation}

Finally, using that $(\Sigma^r - \Sigma^a)_{i,j} = -2 i \Gamma_L 
\delta_{i,1} \delta_{j,1} - 2 i \Gamma_R \delta_{i,N} \delta_{j,N}$,
one can easily check that

\begin{equation}
\theta_F = \frac{1}{2i} \mbox{ln} \mbox{Det} \left[ S \right].
\end{equation}

\begin{figure}
\begin{center}
\leavevmode
\epsfysize=6cm
\epsfbox[300 350  500 800]{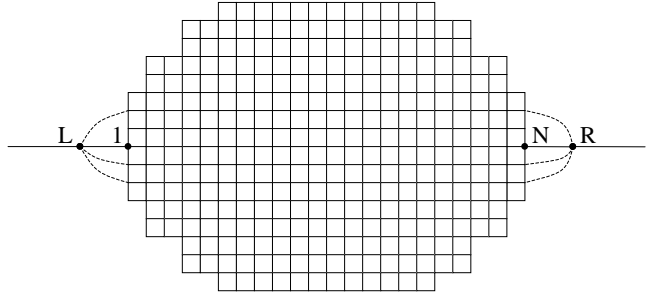} 
\end{center}
\caption[]{Schematic representation of a lattice model for a quantum dot.}
\label{fig1}
\end{figure}

\

\begin{figure}[!th]
\begin{center}
\leavevmode
\epsfysize=7cm
\epsfbox{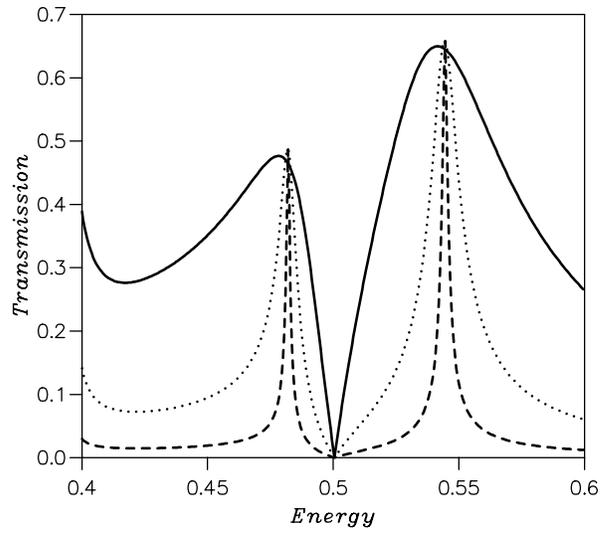}
\end{center}
\caption[]{Transmission probability versus energy for a 5x5 sites dot with
$\Gamma_L = \Gamma_R = 0.25$ (full line), 0.05 (dotted line) and 0.01
(broken line).}
\label{fig2}
\end{figure}

\

\begin{figure}[!th]
\begin{center}
\leavevmode
\epsfysize=7cm
\epsfbox{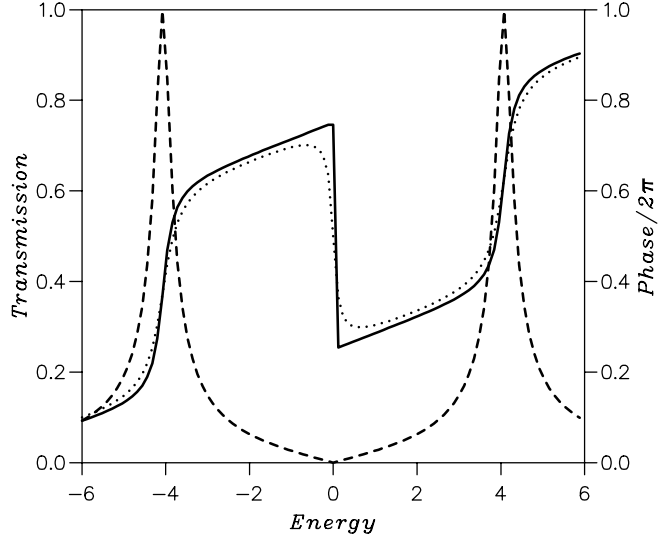}
\end{center}
\caption[]{Modulus (broken line) and phase (full line)
of the transmission amplitude for a four sites dot given by Eq. (18)
with $\Gamma_L = \Gamma_R = 10$ and $\epsilon_2=-\epsilon_3= 4$. The
dotted line shows corresponds to the phase when an inelastic tunneling
rate $\Gamma_{in}=0.1$ is introduced.}
\label{fig3}
\end{figure}

\

\begin{figure}[!th]
\begin{center}
\leavevmode
\epsfysize=10cm
\epsfbox{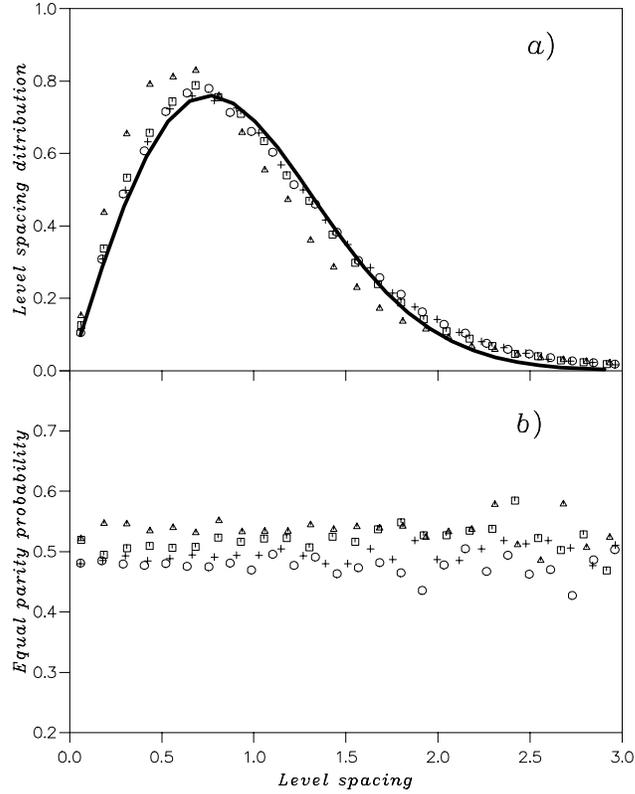}
\end{center}
\caption[]{Level spacing distribution and equal parity probability
for a $30 \times 30$ sites for different values of the disorder strength
$W$: 0.5 (triangles), 1.0 (boxes), 2.0 (crosses) and 3.0 (circles). Full
line corresponds to the Wigner distribution.}
\label{fig4}
\end{figure}

\

\begin{figure}[!th]
\begin{center}
\leavevmode
\epsfysize=7cm
\epsfbox{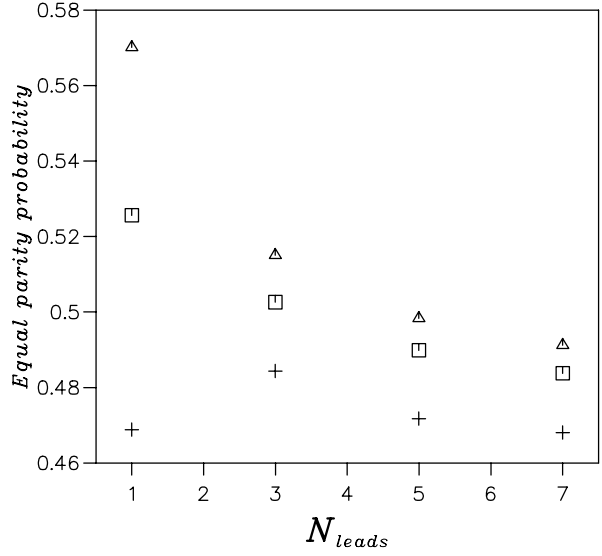}
\end{center}
\caption[]{Equal parity probability
as a function of $N_{leads}$ for
$W = 0.5$ (triangles), $W=1.0$ (boxes) and $W=2.0$ (crosses).}
\label{fig5}
\end{figure}

\

\begin{figure}[!th]
\begin{center}
\leavevmode
\epsfysize=7cm
\epsfbox{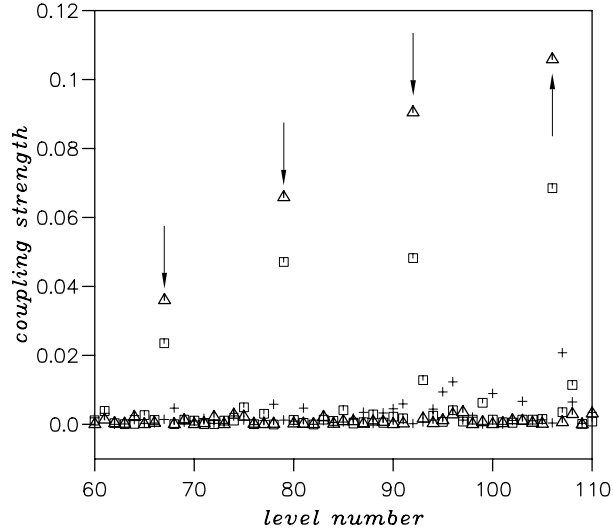}
\end{center}
\caption[]{Strength of the coupling to the leads $\alpha_n$ 
as a function of level
number for the case $N_{leads} = 5$ and different vales of $W$:
0.5 (triangles), 1.0 (boxes) and 2.0 (crosses). The arrows indicate
states with the strongest coupling to the leads for the $W=0.5$ case.} 
\label{fig6}
\end{figure}

\begin{figure}[!th]
\begin{center}
\leavevmode
\epsfysize=12cm
\epsfbox{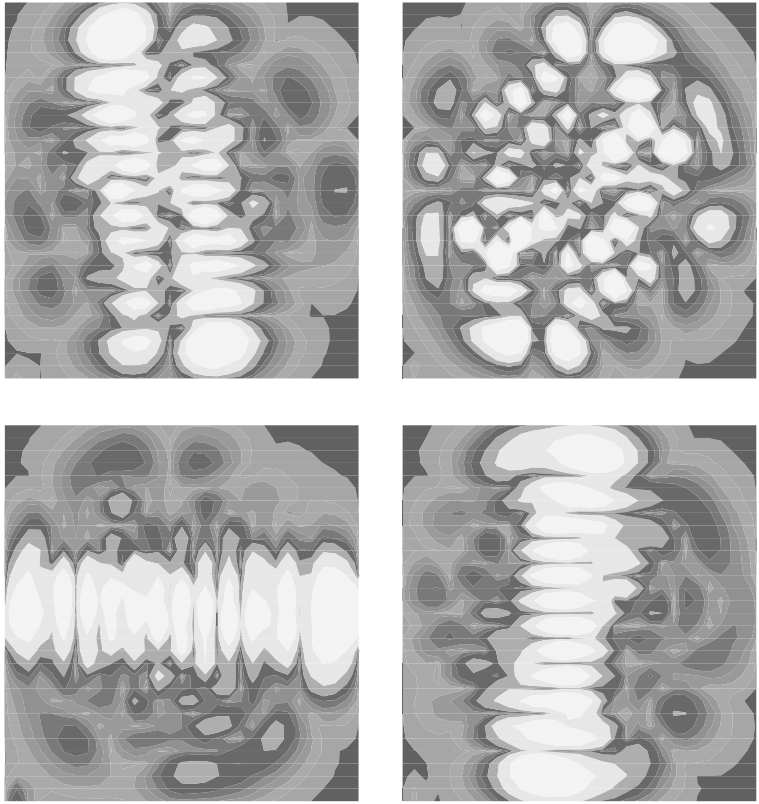}
\end{center}
\caption[]{Contour plots of the wavefunctions for the $W=0.5$ case
corresponding (from bottom to top and from left to right) to level
numbers 67, 68, 69 and 70.}
\label{fig7}
\end{figure}

\newpage

\begin{figure}[!th]
\begin{center}
\leavevmode
\epsfysize=7cm
\epsfbox{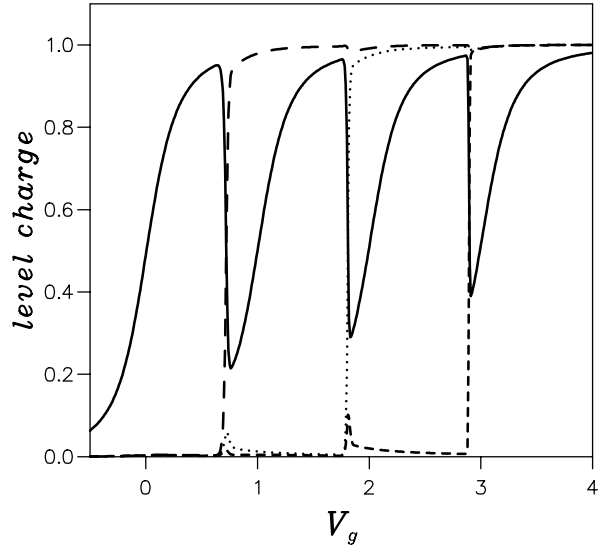}
\end{center}
\caption[]{Level charges as function of gate voltage for a 16 sites
cluster (see parameters in text). The full line corresponds to the 
strongly coupled level and the dashed, dotted and small dashed lines
correspond to the consecutive weakly coupled levels.}
\label{fig8}
\end{figure}

\vspace{1.5cm}

\begin{figure}[!th]
\begin{center}
\leavevmode
\epsfysize=10cm
\epsfbox{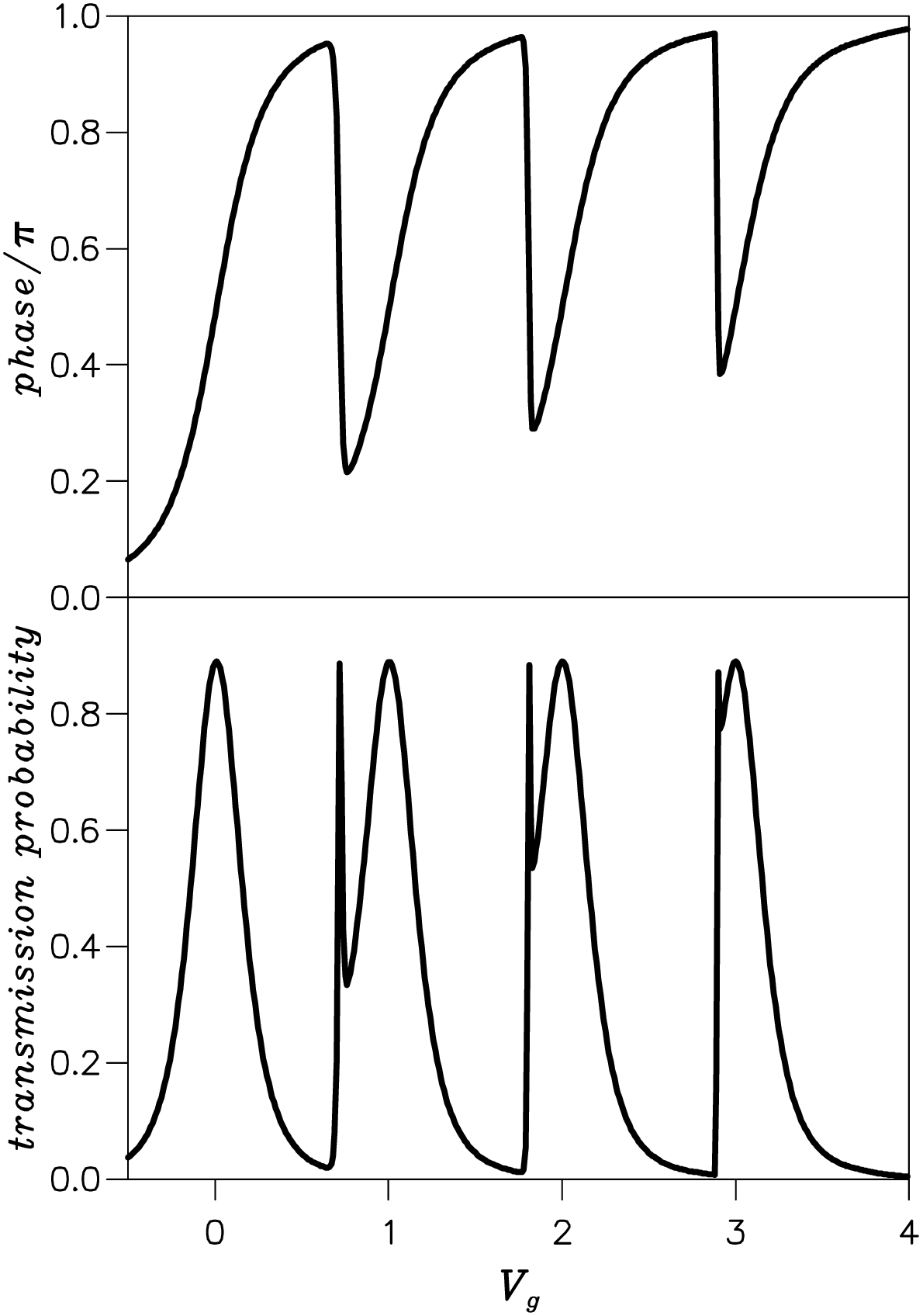}
\end{center}
\caption[]{Phase and modulus of the transmission amplitude for the
model of Fig. 8 mapped into an effective one-electron problem.}
\label{fig9}
\end{figure}


\begin{references}
\bibitem{comment} These results are not dependent on any particular
assumption about the lattice model like band filling, lattice size, etc. 
We thus expect that they remain valid also for a continuous model.
\end{references}
\end{document}